# Temperature stable 1.3 µm emission from GaAs

Slawomir Prucnal,[1,*] Kun Gao,[1, 3] Wolfgang Anwand,[2] Manfred Helm,[1, 3] Wolfgang Skorupa,[1] and Shengqiang Zhou[1]

[1]Institute of Ion Beam Physics and Materials Research, Helmholtz-Zentrum Dresden-Rossendorf (HZDR), P.O. Box 510119, 01314 Dresden, Germany

[2]Institute of Radiation Physics, Helmholtz-Zentrum Dresden-Rossendorf,P.O. Box 51 01 19, 01314 Dresden, Germany

[3]Technische Universität Dresden, 01062 Dresden, Germany

## Abstract

Gallium arsenide has outstanding performance in optical communication devices for light source purposes. Different approaches have been done to realize the luminescence from GaAs matching the transmission window of optical fibers. Here we present the realization of quasi-temperature independent photoluminescence at around 1.3 µm from millisecond-range thermally treated GaAs. It is shown that the $V_{As}$ donor and X acceptor pairs are responsible for the 1.3 µm emission. The influence of the flash-lamp-annealing on the donor-acceptor pair (DAP) formation in the nitrogen and manganese doped and un-doped semi-insulating GaAs wafers were investigated. The concentration of DAP and the 1.3 µm emission can be easily tuned by controlling doping and annealing conditions.

## 1. Introduction

GaAs is one of the most important compound semiconductors widely used for optoelectronic and photovoltaic applications. Much attention has been paid to tune the photoluminescence emission from different GaAs based alloys to the near infrared region (NIR), in particular to the 1.2 – 1.55 µm range, matching the transmission window of silica fibers. It is mainly produced by low temperature molecular beam epitaxy growth of nitrogen or bismuth doped GaAs layers or ion implantation and subsequent annealing [1–7]. The highly mismatched alloys in which the metallic anions (As) are replaced by more electronegative elements (N or Bi) show very strong band gap bowing (>150 meV/% for N and 88 meV/% for Bi) [8]. Recently, the $GaAs_{1-x}Bi_x$/GaAs multiquantum well structures were successfully synthetized by low temperature MBE for developing application in long-wavelength optoelectronics. Such structures reveal both the band gap and photoluminescence intensity weakly temperature dependent [9]. Another approach focuses on incorporating different quantum dots into the GaAs matrix e.g. self-assembled InAs quantum dots (QD) [10]. Depending on the InAs QD size the PL peak emission can be easily tuned from 1.3 to 1.5 µm. Laser diodes operating in the NIR with temperature insensitive oscilation wavelengths are a key issue for telecommunication network systems. Besides manipulation of the optoelectronic properties of the GaAs by doping the proper defect engineering can be a useful method for the photoluminescence enhancement below the band gap of GaAs. The defect related luminescence bands in GaAs are in the spectral range of 0.8 – 1.35 eV and they consist of gallium and/or arsenic vacancy/interstitial complexes [11,12]. The origin of most of them is quite well established in literature besides the 0.95 eV band emission whose the origin is still controversial. Since photons with an energy of 0.95 eV traveling through quartz fibers suffer minimum loss, the GaAs samples containing such luminescence centres can be applied in the field of optical fiber communications.

In this paper we present defect engineering in doped and un-doped GaAs wafers by millisecond range flash lamp annealing [13] for efficient room temperature NIR light emission. The concentration and optical properties of the defects formed in GaAs during thermal processing were investigated by means of positron annihilation spectroscopy, temperature dependent photoluminescence and μ-Raman spectroscopy. It is shown that the $V_{As}$ donor and X acceptor pairs are responsible for the 1.3 μm emission. The highest concentration of the donor-acceptor pairs (the strongest PL signal) was obtained from annealed un-doped and nitrogen doped SI-GaAs wafers. Both types of samples show only 50% reduction of the total PL intensity at 1.3 μm when the temperature rises from 15 up to 300 K. Whereas the incorporation of Mn which is a p-type dopant in GaAs quenches the 1.3 μm PL emission completely due to deactivation of X centers. The influence of the doping type on the optical properties of the SI-GaAs wafers is discussed.

## 2. Experimental setup

Semi-insulating (100) GaAs wafers were implanted with nitrogen and manganese at room temperature. Nitrogen was implanted at energies of 30 and 70 keV with a nominal fluence of $1.1 \times 10^{16}$ ion/cm$^2$ in order to obtain a rectangular depth profile of nitrogen from the surface to 100 nm in depth. The Mn ions were implanted at an energy of 90 keV with a fluence of $2 \times 10^{16}$ ion/cm$^2$. To prevent the decomposition of the GaAs wafers during annealing a SiO$_2$ layer of 200 nm was deposited by PECVD at 200 °C. Implanted and virgin samples were flash lamp annealed for 3 and 20 ms with an energy up to 78 and 155 Jcm$^{-2}$, respectively [13]. The optical and structural properties of annealed samples were investigated by the temperature dependent photoluminescence and μ-Raman spectroscopy. The phonon spectra were determined by μ-Raman spectroscopy in a backscattering geometry in the range of 100 to 700 cm$^{-1}$ using a 532 nm Nd:YAG laser with a liquid nitrogen cooled charge coupled device camera. The same type laser with a power of 60 mW was used for the PL excitation. The PL spectra were recorded at the temperature ranging from 15 up to 300 K using a Jobin Yvon Triax 550 monochromator and a cooled InGaAs detector. The concentration and type of defects were determined by positron annihilation spectroscopy using Doppler broadening using the SPONSOR system at Rossendorf. Positron annihilation peaks were measured using a variable-energy moderated positron beam with energy in the range from 27 eV up to 35.08 keV.

## 3. Results and discussion

The influence of the doping and millisecond flash lamp annealing on the microstructural properties of the GaAs was investigated by means of μ-Raman spectroscopy. Figure 1 shows the first-order μ-Raman spectra obtained from doped and virgin (100) oriented SI-GaAs before and after flash lamp annealing. According to the selection rules in the backscattering geometry the Raman spectra recorded from the (100) oriented monocrystalline GaAs reveals only longitudinal (LO) optical phonon mode at 292 cm$^{-1}$ while the excitation of the transverse (TO) optical phonon mode located at 268.6 cm$^{-1}$ in such geometry is forbidden. In case of virgin samples both non-annealed and flash lamp annealed for 20 ms at 138.7 Jcm$^{-2}$ only the LO phonon mode at 292.2 cm$^{-1}$ is visible. The Raman spectrum obtained from non-annealed nitrogen implanted sample shows two broad peaks at 263 and 285 cm$^{-1}$ corresponding to the TO and LO phonon modes in amorphous GaAs. After FLA at 155 Jcm$^{-2}$ the TO phonon mode disappears and only the narrow LO phonon mode located at 291.2 cm$^{-1}$ is visible. The shift of the LO phonon mode to the lower frequency is caused by GaN$_x$As$_{1-x}$ alloy formation after annealing. Prokofyeva et al. have investigated the influence of the nitrogen concentration on the allowed phonon mode positions in the GaN$_x$As$_{1-x}$ alloys [14]. They have found that the red shift of the LO phonon mode is in the rage of −136 cm$^{-1}$/x in respect to the undoped crystalline GaAs, where x is the nitrogen concentration. Hence, our GaN$_x$As$_{1-x}$ contains about 0.8% of nitrogen which means that 50% of implanted nitrogen was activated. Moreover the nitrogen implanted and annealed samples exhibit a broad peak at around 460 cm$^{-1}$ due to the local vibrational mode associated with nitrogen substituted arsenic sites. The not annealed Mn implanted sample shows two broad TO and LO phonon modes at 260 and 282 cm$^{-1}$, respectively. After FLA at 130.7 Jcm$^{-2}$ both peaks are shifted to the higher wavenumber but still are displaced by 1.0 and 2.5 cm$^{-1}$ to the lower frequency. Due to Mn incorporation into GaAs lattice the annealed samples show the p-type conductivity. According to the selection rules the TO phonon mode in the backscattering geometry from the (100) GaAs is optically forbidden. However, ternary compound semiconductors such as Ga$_{1-x}$In$_x$As show two phonon mode behaviour [15]. The existence of both TO and LO phonon modes in Mn implanted and annealed sample confirm the Ga$_{1-x}$Mn$_x$As alloy formation.

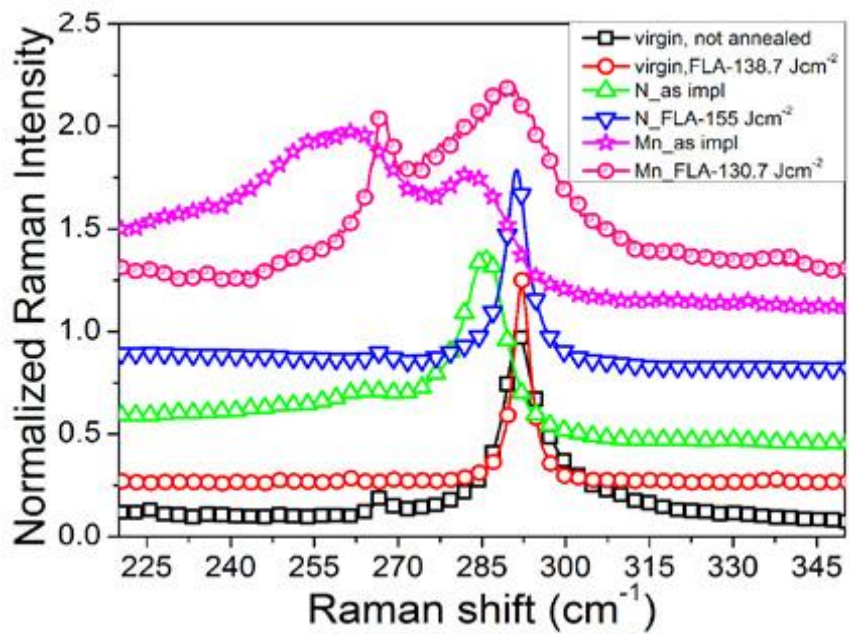

Fig. 1 μ-Raman spectra of implanted and virgin SI-GaAs before and after flash lamp annealing for 20 ms. The spectra have been vertically offset for clarity.

Based on the shift of the LO phonon mode the composition of the Ga$_{1-x}$Mn$_x$As can be calculated according to equation for the strained GaMnAs layer: LO$_{GaAs}$(x) = 292-118.8x, where x is the Mn concentration [16]. In our case the shift of the LO phonon mode is 2.5 cm$^{-1}$ which corresponds to 2% of Mn incorporated into GaAs. Moreover the LO phonon mode is strongly asymmetric and can be de-convoluted into the LO phonon mode and coupled-LO-phonon plasmon mode (CLOPM) usually optically active in p-type heavily doped binary semiconductors [16]. The position of the CLOPM depends on the carrier concentration and moves from the LO to the TO phonon mode position with increasing hole concentration. The Ga$_{0.98}$Mn$_{0.02}$As Raman spectrum fitted with Lorentzian function exhibited the CLOPM at 277.8 cm$^{-1}$ which corresponds to the hole concentration in the range of $1.2 \times 10^{19}$ cm$^{-3}$.

Figure 2 shows the room temperature PL spectra obtained from virgin and N or Mn implanted and flash lamp annealed samples for 3 ms and Mn doped sample annealed for 20 ms. The PL spectrum from virgin not annealed sample is shown as well. For the PL excitation the 532 nm laser with 60 mW power was used. Taking into account the implantation parameters for N and Mn and absorption coefficient α at 532 nm for GaAs [17] the PL spectra of implanted and annealed samples originate only from doped layers. Due to different optical absorption of implanted and not implanted samples the FLA system was calibrated according to the melting point of certain samples in order to obtain the same temperature during FLA process. The maximum activation of the implanted elements appears during liquid phase epitaxial regrowth of amorphized layer. Therefore in each case the annealing was performed at the temperature reaching the melting point of implanted or crystalline GaAs. In order to suppress the decomposition of the GaAs during high temperature annealing a $SiO_2$ layer of 200 nm was deposited on the top. After annealing the oxide layer was chemically etched in $HF:H_2O$ solution before measurements. The room temperature photoluminescence (RTPL) obtained from the virgin not annealed sample shows the near band gap (NBG) emission at 875 nm and weak signal in the near infrared region (NIR). After FLA at 138.7 $Jcm^{-2}$ for 20 ms the strong NIR photoluminescence band at 1.3 μm appears. The same feature of the NIR PL emission reveals nitrogen implanted samples but the NBG emission is observed at 900 nm due to the $GaN_xAs_{1-x}$ alloy formation [18]. The red shift of the NBG emission by 40 meV corresponds to x = 0.007, which is in consistence with the Raman result shown in Fig. 1. The incorporation of manganese into GaAs completely suppressed the NIR PL emission and only the near band gap peak is visible. Moreover manganese forms an acceptor level around 110 meV above the valence band in GaAs [19,20]. It changes the SI-GaAs to p-type material with the carrier concentration in the range of $2 \times 10^{19}$ $cm^{-3}$ for samples annealed at 138.7 $Jcm^{-2}$ for 20 ms. The defects in the virgin and annealed GaAs samples were investigated by positron annihilation spectroscopy (PAS). The virgin GaAs sample shows a linear relationship between the S and W parameter which suggest that only one type of defect exists in this sample which traps positrons. The average positron lifetime ($\tau_{ave}$) was close to the bulk value which is in the range of 230 ps [21]. According to the PAS the high temperature flash lamp annealing in the millisecond range improves significantly the crystallinity of the bulk GaAs. Both values of the S and W parameters after annealing are close to the tabulated values [21]. A small deviation was observed within 400 nm from the surface with S ϵ (0.995 ÷ 1) and W ϵ (1 ÷ 1.05) which is the probed range for the PL excitation. At this range the positron lifetime decreases down to 50 ps. The slight deviation of the S and W parameters from the bulk value and the decrease of the $\tau_{ave}$ indicates the existence of some negatively charged defects coupled to the arsenic vacancy e.g. $V_{As}$-X defect complexes. The arsenic vacancies form shallow donor levels located at about 30, 60 or 140 meV below the conduction band and they are positively charged ($V_{As}^{n+}$) [22,23]. Hence the $V_{As}^{n+}$ alone is not detectable with positrons. $V_{As}$-X defect complexes with negatively charged X defects in the annealed semi-insulated GaAs crystals were identified for the first time by Bondarenko et al. [21]. According to temperature dependent Hall measurements they found that the energy level of the X defects should be located at about 0.5 eV above the valence band and initially was correlated with the $Cu_{Ga}$ acceptor but finally copper was eliminated as the precursor of the X defects. Up to now the origin and the electronic structure of the X-defect in annealed SI-GaAs are unknown. Based on the positron annihilation spectroscopy and photoluminescence results obtained from the undoped and nitrogen implanted SI-GaAs we can conclude that the $V_{As}$-X defect complex is responsible for the 1.3 μm emission where the X defect forms the deep acceptor level about 0.47 eV above the valence band.

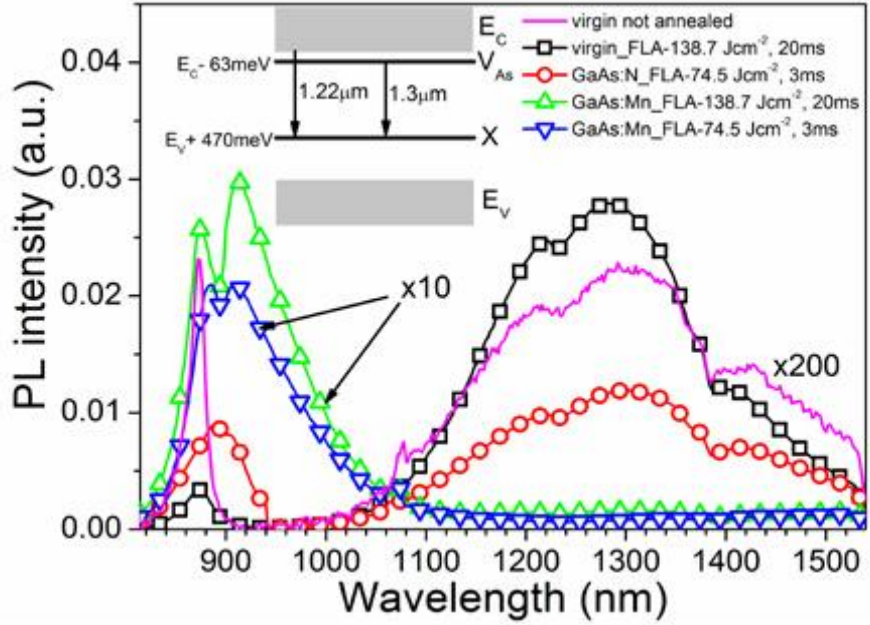

Fig. 2 Room temperature photoluminescence spectra obtained from virgin and implanted SI-GaAs samples after flash lamp annealing. Inset shows the scheme of energy levels and radiative transitions in annealed samples.

In former reports the 1.3 μm emission is frequently related to defect, which occurs in n-doped GaAs [12]. We have tested un-doped and chromium passivated SI-GaAs samples. Independent of the type of SI-GaAs we did not observe any significant changes in the 1.3 μm PL emission besides a small influence on the PL intensity. The absence of the 1.3 μm PL emission in the Mn doped samples confirms our assumption that the negatively charged native defects are responsible for the NIR PL presented in this paper. In the p-type samples such as GaAs:Mn holes neutralize the X- defects and make them optically not active. Besides the 1.3 μm emission the nitrogen implanted and virgin flash lamp annealed samples reveal a PL band at 1.22 μm. The inset of Fig. 2 shows the schema of energy levels of both, the arsenic vacancy and X- defect, with optical transitions responsible for the 1.22 and 1.3 μm emissions, respectively. The $V_{As}$ with an ionization level about 63 meV below the conduction band can be directly populated by optically excited electrons from the valence band [24]. The radiative de-excitation takes place via the X-defects ($E_X = E_V + 470$ meV) with a NIR emission at 1.3 μm. The 1.22 μm emission is due to direct de-excitation of electrons from the conduction band to the X-defect. Different approaches were proposed to tune the NBG emission from GaAs into the NIR region for the optical communication. Most of them are based on ternary alloys formed by introducing other elements such as indium, nitrogen or bismuth into the GaAs crystal matrix. But the proper defect engineering seems to be the simplest method which can be applied for the NIR GaAs light emitter fabrication.

Figure 3 shows the change of the maximum NIR PL intensity as function of temperature after flash lamp annealing. The un-doped sample was annealed at 138.7 Jcm$^{-2}$ for 20 ms while the nitrogen or manganese implanted samples were annealed at 74.5 Jcm$^{-2}$ for 3 ms. In each case the energy deposited to the samples during FLA corresponds to the surface temperature of GaAs of about 1200 °C. The virgin sample shows the strongest NIR PL after FLA for 20 ms while the nitrogen doped sample reveals the highest NIR photon emissivity after annealing at 3 ms. Independent of the annealing time the highest efficiency was obtained after annealing at a temperature close to the melting point of the GaAs. The Mn doped samples show only very weak luminescence in the NIR region and the NIR PL is characterised by strong temperature quenching. Above 150 K the NIR PL signal almost disappears and it is in the detection limit of our PL system. In case of virgin sample the PL intensity of 1.3 μm emission decreases with increasing temperature up to 100 K, then appears thermal population of the $V_{As} - X$ defect complex which reveals as an increase of the NIR PL intensity up to 250 K. Above 250 K the NIR PL starts to decrease again. In the temperature range from 15 up to 300 K NIR PL at 1.3 µm decreases by only 50%. The nitrogen doped sample exhibits monotonic decrease of the NIR PL intensity over the measurement temperature range. The activation energy with respect to thermal quenching significantly increases in comparison with un-doped or Mn implanted samples. The nitrogen doping decreases the band gap of GaAs and changes the conductivity of the implanted layer from semi-insulating to n-type which increases the concentration of photogenerated carriers during photoluminescence excitation. Moreover, nitrogen occupies the arsenic side in GaAs which reduces the concentration of $V_{As} - X$ defect complex and PL intensity at the 1.3 µm. On the other hand N neutralizes the non-radiative crystalline defects centres leading to slow monotonic decrease of the NIR PL over the measured temperature range. The 1.3 µm peak intensity measured at room temperature constitutes 42% of the initial luminescence measured at 15 K. The nitrogen doped samples show slightly higher temperature quenching of the NIR PL compared to the virgin one. The temperature quenching behaviour of the NIR PL emission is the main difference between our results and results existing in literature. The observed NIR PL by other authors from n-type GaAs decreases with temperature by at least three orders of magnitude and was associated with $Si_{Ga}V_{Ga}Si_{Ga}$ or $Ge_{Ga}V_{Ga}Ge_{Ga}$ gallium vacancy-related complexes or recombination luminescence of the DAP, composed of an arsenic and gallium vacancy [12,25,26]. Due to the lack of impurities in the virgin sample we associate the NIR PL with the $V_{As} - X$ defect complex formed in GaAs during FLA.

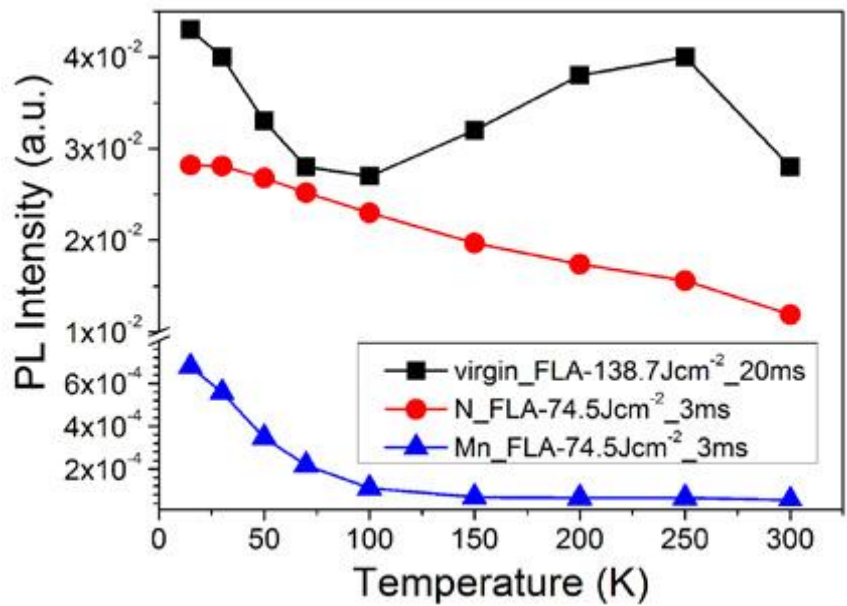

Fig. 3 Maximum PL intensity at 1.3 µm as a function of temperature obtained from virgin and N or Mn doped GaAs after flash lamp annealing.

## 4. Conclusions

In summary, the semi-insulating GaAs samples show promising room temperature NIR PL after flash lamp annealing. The $V_{As} - X$ defect complex is responsible for the 1.22 and 1.3 µm emission. The quasi-temperature stable NIR PL emission can be interesting for potential application in the field of optical-fibre communications. The proper defect engineering presents a simple method for tuning the optical properties of GaAs.

# Acknowledgment

We would like to thank the ion implanter group at HZDR. This work was financially supported by the Helmholtz-Gemeinschaft Deutscher Forschungszentren (HGF-VH-NG-713).

# References and links